# Consent verification monitoring

Marco Robol
  DISI, University of Trento, Trento, Italy, marco.robol@unitn.it

Travis D. Breaux
  Institute of Software Research, Carnegie Mellon University, Pittsburgh, PA, USA, breaux@cs.cmu.edu

Elda Paja
  Computer Science Department, IT University of Copenhagen, Copenhagen, Denmark, elpa@itu.dk

Paolo Giorgini
  DISI, University of Trento, Trento, Italy, paolo.giorgini@unitn.it

## ABSTRACT

Advances in personalization of digital services are driven by low-cost data collection and processing, in addition to the wide variety of third-party frameworks for authentication, storage, and marketing. New privacy regulations, such as the General Data Protection Regulation (GDPR) and the California Consumer Privacy Act (CCPA), increasingly require organizations to explicitly state their data practices in privacy policies. When data practices change, a new version of the policy is released. This can occur a few times a year, when data collection or processing requirements are rapidly changing. Consent evolution raises specific challenges to ensuring GDPR compliance. We propose a formal consent framework to support organizations, data users and data subjects in their understanding of policy evolution under a consent regime that supports both the retroactive and non-retroactive granting and withdrawal of consent. The contributions include: (i) a formal framework to reason about data collection and access under multiple consent granting and revocation scenarios; (ii) a scripting language that implements the consent framework for encoding and executing different scenarios; (iii) five consent evolution use cases that illustrate how organizations would evolve their policies using this framework; and (iv) a scalability evaluation of the reasoning framework. The framework models are used to verify when user consent prevents or detects unauthorized data collection and access. The framework can be integrated into a runtime architecture to monitor policy violations as data practices evolve in real-time. The framework was evaluated using the five use cases and a simulation to measure the framework scalability. The simulation results show that the approach is computationally scalable for use in runtime consent monitoring under a standard model of data collection and access, and practice and policy evolution.

## CCS CONCEPTS

• Security and privacy ~ Software and application security ~ Software security engineering • Security and privacy ~ Human and societal aspects of security and privacy ~ Privacy protections • Security and privacy ~ Formal methods and theory of security ~ Logic and verification • Software and its engineering ~ Software creation and management ~ Software verification and validation ~ Formal software verification • Software and its engineering ~ Software creation and management ~ Designing software ~ Requirements analysis

## KEYWORDS

Privacy, GDPR, Consent, Consent revocation, Retroactivity, Formal framework, Description Logic, Evolution, Verification, Logs, Analysis.

# 1. Introduction

Web and mobile applications increasingly collect sensitive information from application users to offer personalized services. For example, social media and dating websites allow users to build intimate personal profiles that are linked to other users based on personal relationships, employment websites collect and share job histories, and grocery stores use rewards programs to collect and track food and beverage purchases. Personalization allows organizations to tailor user experiences to individual user preferences, and a subset of this data is used to personalize ads, based on changing user behaviors, lifestyle choices and life stages. With cloud-based services, organizations can scale their applications to thousands and millions of users, which requires sharing personal data with third parties [39].

Laws and regulations, such as the EU General Data Protection Regulation[1] (GDPR), have been enacted to transfer control to and from organizations, so that users can restrict how their personal data is collected and used. The GDPR, in particular, protects the data of citizens of EU member countries, while people from other countries remain excluded. While users frequently share their own personal data, the GDPR more generally refer to the *data subject,* which is *the person described by personal data* (see Article 4(1) of GDPR). The term data subject is sufficiently broad to cover data processed by first-party services, who have a direct relationship with users, and data processed by third-party services who have an indirect relationship with users, as well as to account for data shared by one user that is about another person. Prior work to formalize consent under the GDPR [6, 17] have considered the logical consequences of data collection, access, and purpose when consent is granted or revoked. However, as policies and data practices change, these formalizations can become inconsistent. This paper addresses this limitation of prior work.

Under the GDPR, consent is a key element in privacy, and it has become a critical element under the EU General Data Protection Regulation (GDPR[1]). Under GDPR, consent is one of a few legal bases available that can be used to process app user data (see Article 6-1(a)) and, in most cases, consent is the only viable basis. Consent constitutes legal evidence of app user awareness about their data being collected, used, and shared by organizations. Under the GDPR, the data subject is protected because the demonstration of a valid acquisition of the consent is a responsibility of the organization, i.e., the data subject does not need to initiate a request to receive this protection. To this end, the organization must present information about how the data will be processed, and then request consent from the data subject before processing the data. Furthermore, the data subject may revoke their consent at a later date, which means that the organization can no longer process data collected after the revocation (see Article 7). However, the organization may continue to process data collected under the previously granted consent, if they choose to do so. In general, the organization or data user often obtains consent through click-through privacy notices, e.g., when the app user first uses the service. Consent can be obtained in other ways, including just-in-time consent after an app user has already begun using the service, as long as the consent is granted prior to the collection or use of the subject's data [42].

Data subject decisions about granting consent can be driven by the perception of trust in an organization with respect to their history of bad privacy practices. For example, recent disclosures by Facebook of personal data leaks to third-parties [22] led some app users to restrict their privacy preferences, which control who can access their data, and others to delete their account, thus opting out entirely.

Internally, organizations can make changes to their data practices several times a year. Evidence of changes can be observed in the revision histories of evolving privacy policies. Figure 1 shows the changes to the privacy policy of Waze, a popular mobile app for automobile navigation, based on an analysis by the authors [38]: the y-axis shows the number of statements per policy revision, with the policy revision dates along the x-axis; exact statement matches appear in blue, new statements appear in red, and statements with changes to wording appear in orange. Some of these changes are due to changes in boilerplate language (e.g., how the organization or user are referenced), or to data purposes. Under the GDPR, changes to data practices and purposes require consent. While Waze in particular underwent a number of changes from late 2012 to mid-2014, there were significant changes from late 2017 through mid-2018, at which point the GDPR went into effect.

---

[1] Regulation (EU) 2016/679 of the European Parliament and of the Council of 27 April 2016 on the protection of natural persons with regard to the processing of personal data and on the free movement of such data, and repealing Directive 95/46/EC (General Data Protection Regulation)

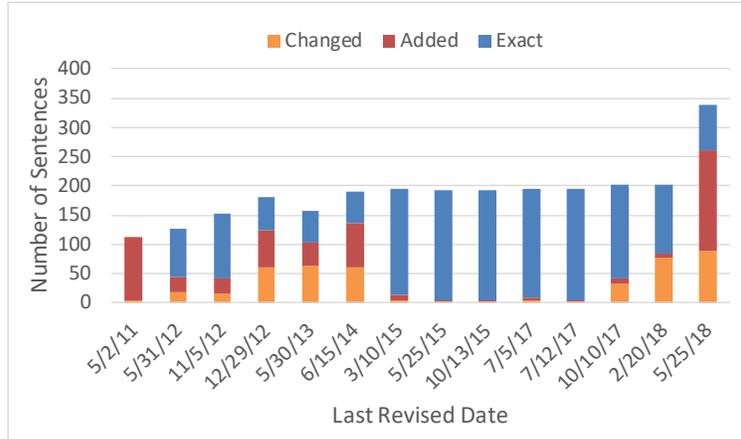

**Figure 1: Revision History of the Waze Privacy Policy from 2011-2018**

Changes in an app's data practices can be driven by changes in their underlying technologies. For example, as a mobile navigation app, Waze accesses a user's real-time location using the mobile device's operating system to offer real-time navigation and driving-related services. Location data can include more information than simply the user's longitude and latitude position. Depending on the technology used to calculate the position, whether the user is indoors, outdoors, walking or driving could all be inferred, and if the location were linked to a location database, whether the user was located in a residential or retail space and even which retail store could be inferred. Moreover, if the position is calculated with the use of short distance technologies, such as Bluetooth or 802.11 wireless networking, information about other devices may be collected, even if not pertinent to the purposes of the app. User movements over time could be stored to suggest frequently used routes, visited locations and preferred retail stores. The evolution of location technologies generates different location data types, wherein precision, frequency, and other tracking information differs from one technology to another. The varying sensitivity of types of data collected and accessed are, in part, driven by the evolution of the technology. In parallel, as technology drives changes in data practices, an organization must update their privacy policies and ensure that previously granted consent matches the scope of data collected and its purposes of use.

As a matter of requirements engineering, under the GDRP see Articles 6-1(a) and 7, organizations should tag their data to know when it was collected and when they obtained consent. Because the GDRP requires that consent be granular, including that purposes be distinguishable (see Recitals 32 and 43), organizations should also tag this data with purposes for which consent was granted. Notably, organizations may collect data as a consequence of their system design, but they may not process the data for a specific purpose without consent. At scale, one can imagine that organizations who are in competitive markets will be looking for new opportunities to process app user data, leading to changes to their practices. In addition, app users may either be uncomfortable with new purposes, or shift their trust in organizations due to improper data handling by the organization or the market. To address this problem, we propose a formal consent framework, expressed in Description Logic (DL), to support data subjects, data users and organizations in the understanding of consent under evolving policies. The framework formally expresses how to automatically verify the compliance of data access, given data subject consent. The contributions of the paper include: (i) a formal framework that can be used by organizations to support data users to reason about data collection and access under multiple consent granting and revocation scenarios; (ii) a scripting language that implements the consent framework for encoding and executing different scenarios; (iii) five consent evolution use cases that illustrate how organizations would evolve their policies using this framework; and (iv) a scalability evaluation of the reasoning framework to show the computational feasibility of the approach.

These three contributions above comprise new, previously unpublished work that extends a "new ideas" paper originally presented in RE'19 in track RE@Next! [38].

## 2. Framework Overview and Terminology

The formal consent framework is expressed in Description Logic and designed for organizations to reason about whether data collection and access are authorized by a data subject's consent history, given that consent can be granted and withdrawn under one of two modalities: *retroactivity*, in which the act of granting and withdrawing consent are applied retroactively to affect access to previously collected data; and *non-retroactivity*, in which granting and withdrawing consent only affect access to data in the present and future.

In this framework, we distinguish between the web and mobile *application user*, who is often the data subject about whom data is collected, and the *data user*, who works for an organization or organization and uses the data subject's data for business purposes. The data user is a general term that covers the data processor role under Article 4(2) and (6) of the GDPR, including collecting, recording, storage, alteration, retrieval, use, transmission and erasure, among other activities. These activities are generalized into two principal actions: *collection*, which describes the point where data enters the organization, and *access*, which covers activities to use the data for a specific purpose. Activities, such as erasure and destruction, are outside the scope of the current framework. In Section 6, we discuss how the "Right to Be Forgotten" is separate from the consent framework. In this paper, *policy* refers to the abstract rules that govern an organization's data practices. These policies may be summarized in natural language in the form of a "privacy policy," which is a legal document organizations use to communicate their policy and practices to data subjects and users. The framework described in Section 4 is a formal representation of a specific subset of the organizations overall policy intended to govern the matter of consent to collect and access personal data. Hypothetical or real changes to the policy can be expressed in the framework to reason about the effects of those changes, and laws require that privacy policies are updated to accurately reflect the organization's policy, whenever it changes. The framework can be used to verify polices and data practice scenarios at design time, or it can be integrated into a runtime architecture to perform monitoring.

Figure 2 shows an overall system architecture in which the consent framework could be used to monitor consent at runtime. In Figure 2, we distinguish between authorized services, which go through an access control gateway in order to access the database, and trusted-by-design services, which directly access the database, but are still monitored via a logging service. The architecture represents how consent can be verified and monitored as follows: (A) when users grant and withdraw consent, a consent service records those changes to a consent log via a logging service. The consent log is used by the consent monitor to update the consent model. In addition, (B) services that are trusted-by-design will access a database, which (C) logs queries to an access log that is (E) monitored by the consent monitor to detect ex post violations. Moreover, (F) services that require per-transaction authorization will request permission using an access control service (ACS), which can either (F) log those requests to an access log for ex post violation detection, or (G) the ACS can delegate those requests to the consent monitor, which would deny a request prior to access, if a valid consent permitting access has not been granted.

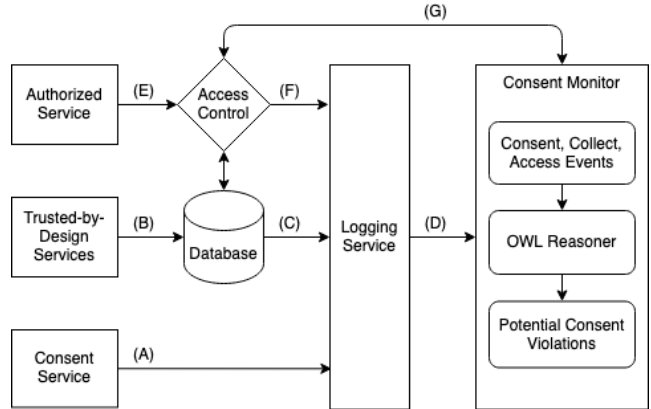

**Figure 2: Architecture to Illustrate Monitoring**

In Figure 3, we present a consent framework example to illustrate the overall approach and consent lifecycle. The example includes principal components of the framework model and language, including actor actions (grant and revoke consent, collect and access data). These components are formally defined in Section 4. In the figure, the actions performed by the data user and data subject are shown as "swim lanes" where actions, represented by ovals, are presented in time order, indicated by solid black arrows that lead from past events to future events. At each point where a data user collects or accesses data, the consent framework can be used to check if the action is approved by the data subject's consent history, including where consent was granted and withdrawn, or if the action is denied. In Figure 3, we illustrate where a non-retroactive withdrawal of consent prohibits a future collection (see red "access denied" sign in the Data User swim lane), but does not prohibit access to a data previously collected under the withdrawn consent (see subsequent "access approved" sign in the same swim lane). This example is simple, and more complex scenarios include overlapping consents and other modes of retroactive and non-retroactive consent and withdrawal.

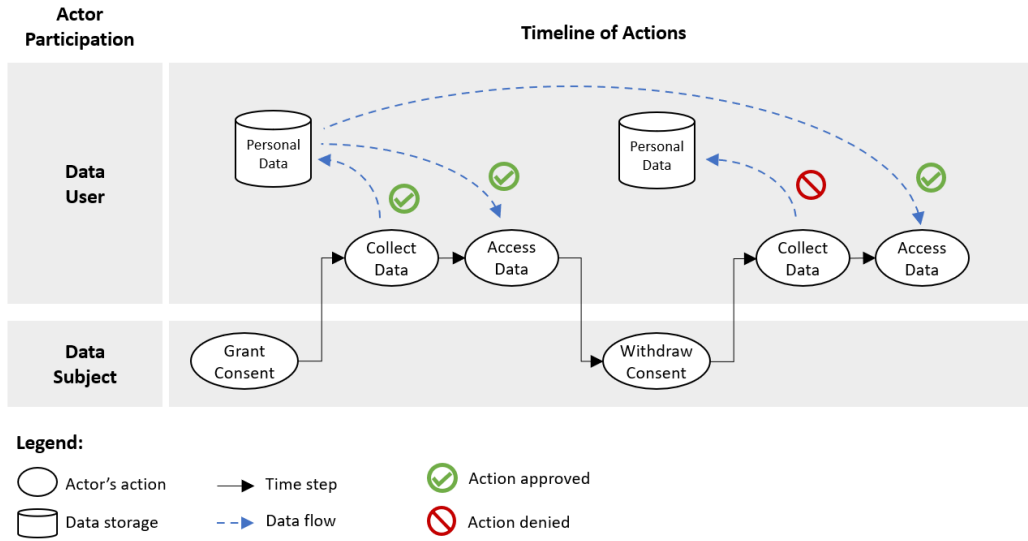

**Figure 3. Example of consent framework illustrating a simple scenario within the framework**

Under the GDPR, organizations are permitted to use retroactive consent and non-retroactive withdrawal. There are no restrictions in GDPR preventing organizations from using non-retroactive consent or retroactive withdrawal, which both provide data subjects with greater privacy protection, but which would make less data available to data users. Figure 4 presents six scenarios that arise from the combination of consent and withdrawal and both retroactivity modalities. In Figure 4, time moves forward from left to right, and horizontal shaded bars show current access authorizations by collection time. Dark green shading shows collection time from which data is accessible; the light red shading shows where data is inaccessible. The vertical lines show times where consent is granted or withdrawn. Non-retroactive consent (top, scenario 1) grants access only to data collected in the future. While non-retroactive withdrawal (scenario 2 and 5 from top) revokes access to data collected in the future, but not to previously authorized data access. Finally, retroactive withdrawal (scenario 3 and 6) revokes access to data at any collection time, while retroactive consent (scenario 4) grants access to data at any collection time.

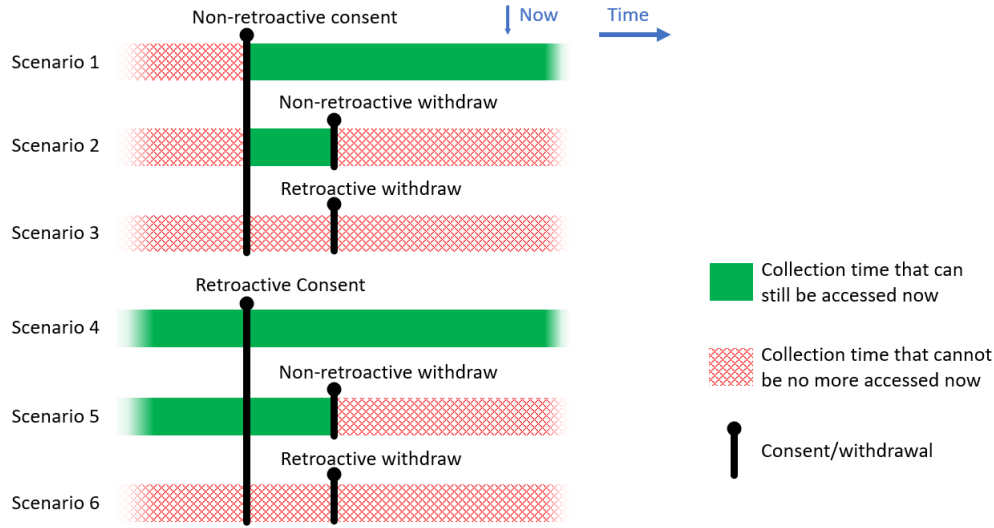

**Figure 4: Six scenarios combining retroactive and non-retroactive consent and withdrawal**

The consent framework is formalized in Description Logic to allow a data user to express and check an arbitrary number of consent and withdrawal events while ensuring compliance at collection and access time. We next discuss the technical challenges of formalizing the framework, before presenting the framework and illustrating examples.

Organizations can apply the framework by formalizing their data practices in the limited scope of data collection and access. This formalization includes constructing an ontology of data types and data purposes, which must be maintained by the organization over time as their software and associated technologies evolve. Using the framework, the organization can then (1) simulate consent evolutions to understand the effect of using any of the six consent scenarios (see Figure 4); (2) they can adopt any one of the six scenarios as a reference model in the design of GDPR-compliant software; or (3) they can monitor and verify at runtime whether a request to collect or access data is approved or denied based on the consent history of the data subject (see Figure 3). The consent framework, whether applied in a simulation or in consent verification, supports an organization who processes data using any combination of the six scenarios applied to the same data types.

## 2.1 Consent and Right to be Forgotten

Consent under the GDPR can be withdrawn at any time by the user. When this happens, organizations must quickly halt the processing of user data and block any further access to such data. Data collected under previously granted consent may still be accessed if consent is withdrawn non-retroactively, or if there is another valid legal basis or another consent that covers that same data and has not been withdrawn.

The GPDR provides the right to be forgotten, which is similar to withdrawing consent at any time. Recall that retroactive withdrawal is a stronger protection than non-retroactive withdrawal, because retroactive withdrawal removes authorizations to access data previously collected. This effectively means the previously collected data can still reside in an organizations database, but it is no longer accessible under the consent that was withdrawn. With retroactive consent granted in the future, the data may become accessible once again.

The right to be forgotten is a stronger protection than retroactive withdrawal, because the users can now ask for their data to be deleted, in addition to halting all processing activities. Organizations must delete all user data and "forget" information related to this user. In the case of data collected under a different legal basis than consent, for example, and in the case of public information whose collection did not require an explicit consent of the user, the right to be forgotten can also be used to delete that data. If data is accessed, used and shared for multiple purposes, all these activities must stop and the same is true for third parties, who are asked to delete the data and stop further data processing activities.

Next, we introduce the technical challenges to consent verification, before we introduce the framework formally in Section 4, and describe framework application use cases in Section 5.

## 3. Technical Challenges

Description Logic (DL) [10] are the de-facto languages for ontologies and the Semantic Web. DL is a subset of first-order logic languages, which is less expressive but guarantees decidability. We chose DL because of its ontology orientation, since we need to represent and verify hierarchical relationships between the data types used in privacy policies. For example, users can provide consent on some broad category of data, such as *personal information*, or they can provide separate consents on narrower, more specific categories, such as *e-mail address* or *phone number*. These hierarchical relationships can lead to inconsistencies and conflicts in deciding if data can be processed [18].

While DL can be used to more easily build consistent taxonomies from the same kinds of data types used in privacy policies, the formalization of consent evolution is more challenging in DL due to their limited expressivity and an open world assumption. DL has limited expressivity, that has been introduced in favor of decidability [31]. DL supports only binary relations, so that two individuals can be related one to another, but no other dimensions, such as time, can be included within the same relationship.

The open world assumption means that unknown facts are considered neither false nor true. In contrast, a closed world assumption, also called negation by failure, considers unknown facts to be false by default. In DL, if one declares that "approved consent" is equivalent to "not consent withdrawn," then consent events can still be in an unknown state in which they are neither consented nor withdrawn.

**Monotonicity**. Monotonicity is a desirable property for updating functions of any formalization, because additions to a knowledge base with this property will not invalidate prior facts. Monotonic update functions are simpler and efficient, and ensure that the existing knowledge base is never changed and only extended by new facts. However, consent evolution appears to have a non-monotonic behavior. For example, in the case of a withdrawal event, one might expect the update function should change existent authorizations in a non-monotonic way by negating accessibility, or changing the authorization from a permission to a prohibition. While consent may be implemented using various access control rules across different systems and platforms, consent only defines a window of authorization between acts of granting and withdrawal consent, and not specific access control rules. Within this window, other concerns may affect who has access based on fine-grained access control rules subject to platform implementations. To ensure monotonicity, we formalize the evolution of these authorization windows in our representation, so that the whole history of consent and authorization changes is maintained. We use temporal steps to define the validity of authorizations over time, specifically, consent is withdrawn from the temporal step in which the authorization is no longer valid. In contrast, consent is approved at the temporal step in which authorizations start to be valid. This distinction will be illustrated further later in Section 4.1.

**Temporality**. Representation of time and temporal concepts is not specifically supported in DL. There are temporal extensions for DL [7, 8, 9, 32], but they all introduce overwhelming extensions of the language and an increase in computational complexity, which we aim to avoid in favor of understandability and efficiency.

For example, some temporal extensions are based on the representation of many time-related concepts, relationships, and constraints, such as an *instant* or *interval* of time, the concepts of *before*, *after*, *meanwhile*, *started before*, *ended after*, and so on. Moreover, the representation of temporal concepts, such as the interval of time in which a consent was granted and then withdrawn, can be approached in different ways. A strong simplification of the representation of temporality is to focus only on forward-time or backward-time. In our framework, we have multiple, evolving consents, where a change in consent affects authorizations only in the future and never in the past. Thus, our temporality representation is based on forward-time.

**Use Limitation**. The GDPR requires organizations to limit uses to those purposes stated at the time of collection, which is also called the Use Limitation Principle in the OECD Guidelines on the Protection of Privacy and Transborder Flows of Personal Data. To implement use limitation, organizations must specify purposes in advance and restrict access to collection purposes. Purposes may be broadly described, such as advertising, or they may be specific, such as targeted advertising or payment processing. In Role-Based Access Control (RBAC), roles represent a "competency to do specific tasks" or "duty assignments" to data users, which change infrequently because they correspond to an organization's functions and business processes [40]. The key difference between roles and purposes is the orientation: roles describe the work performed by a class of users, whereas purposes define the work for which data is used. For example, one can assign the same scope of work to the role Advertiser as they would the data purpose Advertising. Because of the extensive history of using role-based orientations in established security standards, we use the role-based orientation in this paper to implement use limitation, noting that a purpose-based orientation could alternatively be used, if preferred.

## 4. Formal Consent Framework

This section presents the formal consent framework. The formal framework is specified in Description Logic (DL), which is a subset of first-order logic for expressing knowledge. A DL knowledge base KB is comprised of intensional knowledge, which consists of concepts and roles (terminology) in the TBox T, and extensional knowledge, which consists of properties, objects and individuals (assertions) in the ABox A [10]. In this paper, we use the DL family *ALC*, which includes logical constructors for union, intersection, negation, and full existential qualifiers over roles. Concept satisfiability, concept subsumption and ABox consistency in ALC are PSPACE-complete [10]. In the notation that we use in this paper, strings of lowercase letters are used for individuals and strings with the initial letter as an uppercase letter are used for concepts.

Description Logic includes axioms for subsumption, disjointness, and equivalence with respect to a TBox, and an interpretation function $\cdot^{\mathfrak{I}}$ that assigns an interpretation to a concept. The special concepts "top" $\top$ correspond to the entire domain of interpretation $\top = \Delta^{\mathfrak{I}}$, and the "bottom" concept $\bot$ corresponds to an empty domain $\bot = \emptyset$. Subsumption describes individuals using generalities: we say a concept C subsumes a concept D, written $T \vDash D \sqsubseteq C$, if $D^{\mathfrak{I}} \subseteq C^{\mathfrak{I}}$ for all interpretations $\mathfrak{I}$ that satisfy the TBox T. The concept C is disjoint from a concept D, written $T \vDash D \sqcap C \to \bot$, if $D^{\mathfrak{I}} \cap C^{\mathfrak{I}} = \emptyset$ for all interpretations $\mathfrak{I}$ that satisfy the TBox T. Finally, the concept C is equivalent to a concept D, written $T \vDash D \equiv C$, if $D^{\mathfrak{I}} = C^{\mathfrak{I}}$ for all interpretations $\mathfrak{I}$ that satisfy the TBox T.

The universe of discourse consists of concepts for $Collection$, $Access$ and $Consent$ with associated events described by individuals, e.g., a collection event $c \in Collection$, and concepts for $Data$, $DataSubject$, the $Recipient$ of data during an access event, and $Time$. The consent history is a DL knowledge base KB that is monotonically constructed from a series of collection, access and consent events. We now describe definitions needed to perform this construction. In this formalization, we make a few key assumptions underpinning the consent history:

- Updates to the consent history are monotonic, i.e., new information cannot invalidate previously received information. This assumption is imposed by Description Logic, which is monotonic.

- When granting and withdrawing consent, authorizations describe data in vague or abstract terms. This is consistent with modern practice, where organizations seek authorization for broad classes of data to increase flexibility in data practices.
- When collecting or accessing data, organizations describe data in specific terms. This is consistent with how hardware, software and algorithms are written to operate on specific data elements, or collections of specific elements.

## 4.1 Representing Time

The consent framework uses the notion of forward time, in which events can be referenced as occurring after other events, but there is no way to express the occurrence of events during or before other events. While limited, this restriction is both necessary and minimal.

**Definition 1**. *Time* is described by subsumption, using the notion of *forward time*, such that it is true that $T \vDash T_x \sqsubseteq Time$ with respect to a TBox T, and for each time concept $T_x$ that represents an open time interval beginning at a time step $x$ in $1\ldots n$ up to the latest time step $n$. Each time step $x+1$ in the future is described by the concept $T_{x+1}$, such that it is true that $T \vDash T_{x+1} \sqsubseteq T_x$.

The linear chain of subsumption in which nested concepts contain exactly one concept per level provides a complete description of time, with a root concept $Time$ that includes any time since the beginning, a leaf concept $T_n$ that includes only time in the future, and no disjoint concepts. Thus, in forward time, each concept $T_x$ describes a time interval that begins at time step $x$ and continues forward indefinitely. Concepts that represent time starting from a given instant in the future, are subsumed by concepts that represent time starting from an instant in the past.

Events can be placed within time intervals explicitly as individuals in Description Logic. Time intervals are described by the conjunction of a start time, inclusive, and an end time, exclusive. The end time is represented by the negation of Time at the end time step. The interval $T_x \sqcap \neg T_y$ for $y > x$ describes events that occur in times $T_x, T_{x+1}, T_{x+2}, \ldots$ up to and not including time $T_y$ and beyond. Under the open world assumption, explicit negation is used to close the time interval, otherwise the event can be subsumed by $T_x$ and any subclasses in $T_x$.

Forward time is a restricted representation of time in logic, which is sufficient and minimal to represent time for collection, access and consent.

## 4.2 Representing Data

In the consent framework, data is the principle concept about which events take place. Data concepts may be broad and encompassing, such as personal information, or specific, such as e-mail address [26]. Descriptions of data are broad when referenced in consent events, so as to support authorization to collect and access a class of data, whereas they are specific when referenced by collection and access events for which a specific data type is sought for a specific purpose.

**Definition 2**. *Data* is described by subsumption, such that it is true that $T \vDash C \sqsubseteq Data$ for each data concept $C$ with respect to a TBox T. Organizations can introduce an arbitrary number of data concepts, and should declare when it is known that two data concepts are disjoint. For example, an organization may declare that $Age$ and $Location$ are disjoint, but not declare that $PhoneNumber$ is disjoint from $DemographicData$, if this separation is unknown.

In the consent history, interpretations of Data are constructed through relations, called properties, to other concepts in the knowledge base. These interpretations answer specific questions, such as "which data subject is the data about," or "when was the data collected?" Each property narrows the interpretation of data in question, e.g., given a data subject $datasubject1 \in DataSubject$, the axiom $about(datasubject1) \sqsubseteq$ $Data$ refers to all data about this data subject, including data that has been collected and accessed. The framework formalization supports five properties to data described in the consent history:

- $Data \sqcap \exists about.DataSubject$ – the *about* property is asymmetric and non-transitive and describes who the data is about.
- $Data \sqcap \exists collectedAt.Time$ – the *collectedAt* property is asymmetric and non-transitive and describes when (within what time interval) the data is collected.
- $Data \sqcap \exists collectedBy.Recipient$ – the *collectedBy* property is asymmetric and non-transitive and describes by whom the data is collected. The *Recipient* concept and its subclasses describe the job role or purpose for which data is collected.
- $Data \sqcap \exists accessedAt.Time$ – the *accessedAt* property is asymmetric and non-transitive and describes when (within what time interval) the data is accessed.
- $Data \sqcap \exists accessedBy.Recipient$ – the *accessedBy* property is asymmetric and non-transitive and describes by whom the data is accessed. The *Recipient* concept and its subclasses describe the job role or purpose for which data is collected.
- $Data \sqcap \exists authorizedBy.Consent$ – the *authorizedBy* property is asymmetric and non-transitive and describes which subclass of data is authorized by a consent for collection and access. This property has the inverse property *authorizes*.

The above properties will be used below to define how events are recorded in the consent history.

### 4.3 Collection, Access and Purpose

Consent to data collection and access should be restricted by data purpose, which should be granular and separable under GDPR. Data purpose granularity refers to the level of specificity of the purpose description: a purpose "for service improvement" is less granular than "for identifying and fixing software faults" and "for improving response times," which are more granular and examples of service improvement activities. Data purpose is needed to apply the use limitation principle, which guarantees that data is only used for the original purposes for which the data was collected, and for no other purposes. In the consent framework, purpose is expressed using a role-based orientation, in which the data user assumes a role that describes the class of work for which the data will be used. For example, a data user may be an advertiser or payment processor, and moreover, we assume that any data-related class of actions can take on a role-based orientation. In the consent framework, all data user roles are a sub-class of the *Recipient* class.

Whenever data is collected or accessed, a new event is recorded in the consent history and represented as an individual in the ABox. Each event has properties that answer questions about when the data was collected or accessed, who collected or accessed the data, and who the data is about.

**Definition 3**: *Collections* are events with a relation to the time of collection and the recipient of the collection, as expressed by the DL intersection $Collection \sqcap \exists collectedAt.Time \sqcap \exists collectedBy.Recipient$. For a data type $Location$, a data subject $datasubject1$, and a recipient $Advertiser$, we express a collection event $c1$ at time $T_1$ as follows:

$c1 \in Collection \sqcap \exists collectedAt.(T_1 \sqcap \neg T_2) \sqcap \exists collectedBy.Advertiser \sqcap about.datasubject1$

Where $T_2$ is the closest time step following $T_1$, so that $T_1 \sqcap \neg T_2$ limits the collection between $T_1$ and $T_2$.

**Definition 4**. Accesses are events with a relation to the time of access and the recipient of the access, as expressed by the DL intersection $Access \sqcap \exists accessedAt.Time \sqcap \exists accessedBy.Recipient$. For example, an access event $a1$ at time $T_1$ is expressed as:

$a1 \in Access \sqcap \exists accessedAt.(T_1 \sqcap \neg T_2) \sqcap \exists accessedBy.Advertiser \sqcap about.datasubject1$

To ensure a consistent consent history, however, each collection and access event must be authorized, which we now discuss.

## 4.4 Granting and Withdrawing Consent

Granting and withdrawing consent is an action performed by the data subject. These events define the space of authorized collection and access over a time interval, which is opened at the time of granting consent, and closed at the time of withdrawal.

**Definition 5**. *Granting consent* is an event with an authorization relation to: (1) an entire class of data; (2) any acts to collect or access this data class at an open time interval; (3) the recipient by whom the class was collected; and (4) the subject about whom the data describes.

In the consent history, each consent event is represented by an individual in the ABox. When consent is granted non-retroactively, the consent denotes an authorized interpretation of data collection and access.

We express a consent $consent1$ to collect and access location data by an advertiser from time $T_2$ onward, as follows:

$$consent1 \in Consent \sqcap \forall authorizes. Location$$

$$\sqcap \; \forall authorizes. \begin{pmatrix} (Collection \sqcap \exists collectedAt. T_2) \\ \sqcup \; (Access \sqcap \exists accessedAt. T_2 \sqcap \exists collectedAt. T_2) \end{pmatrix}$$

$$\sqcap \; \forall authorizes. (\exists collectedBy. Advertiser \sqcap about(datasubject1))$$

In the above axiom, the collection time constrains data in both collection and access, whereas the access time only constrains data in access. This distinction is because consent may be granted *retroactively*, which means that, while data may only be accessed from time $T_2$ onward, this access may be to data that was collected before time $T_2$. The retroactive variant of the above consent is expressed as follows, where the concept $Time$ represents the earliest beginning of time:

$$consent1 \in Consent \sqcap \forall authorizes. Location$$

$$\sqcap \; \forall authorizes. \begin{pmatrix} (Collection \sqcap \exists collectedAt. T_2) \\ \sqcup \; (Access \sqcap \exists accessedAt. T_2 \sqcap \exists collectedAt. Time) \end{pmatrix}$$

$$\sqcap \; \forall authorizes. (\exists collectedBy. Advertiser \sqcap about. datasubject1)$$

**Definition 6**. *Withdrawing consent* is a DL extension to an existing, unwithdrawn consent event by closing the previously open time interval for collection or access through the authorization relationship. For example, to withdraw consent, we simply update the class description of the consent individual. Non-retroactive withdrawal at time $T_4$ is expressed as follows:

$$consent1 \in \neg \forall authorizes. \begin{pmatrix} (Collection \sqcap \exists collectedAt. T_4) \\ \sqcup \; (Access \sqcap \exists collectedAt. T_4) \end{pmatrix}$$

Unlike non-retroactive withdrawal that continues to permit access to data previously collected in the authorized time interval (e.g., from $T_1$ to $T_4$), retroactive withdrawal prohibits future access to previously collected data. The retroactive variant of the above withdrawal is expressed as follows:

$$consent1 \in \neg \forall authorizes. \begin{pmatrix} (Collection \sqcap \exists collectedAt. T_4) \\ \sqcup \; (Access \sqcap \exists accessedAt. T_4 \sqcap \exists collectedAt. T_4) \end{pmatrix}$$

As described in Section 4.1, time is represented using forward time. In addition, actions to access data may not occur before that data is collected. These two restrictions have important consequences on how we reason about consent. In Figures 5 through 8, we present four visualizations for consent that examines each combination of non-retroactively and retroactively granted and withdrawn consent over an arbitrary data concept. Access can occur at the same or future time of a corresponding collection over that data. The blue lines indicate the start of a time interval, and the red lines indicate the end of the time interval. Collection and access authorized by each consent type are shown as circled check marks distributed over times $T_1, T_2, \ldots$

In Figure 5 to 8, collection time is represented vertically, and access time is represented horizontally. Blue and red lines delimit the accessibility of the data with respect to collection time. More precisely, blue lines represent the time when consent was granted, while red lines represent the time when consent was withdrawn. The horizontal disposition of the lines corresponds to non-retroactivity, for which access is restricted to a given collection time. The vertical disposition of the lines corresponds to retroactivity, for which access is restricted only between the times of consent and withdrawal with no limitation on the original collection time of the accessed data.

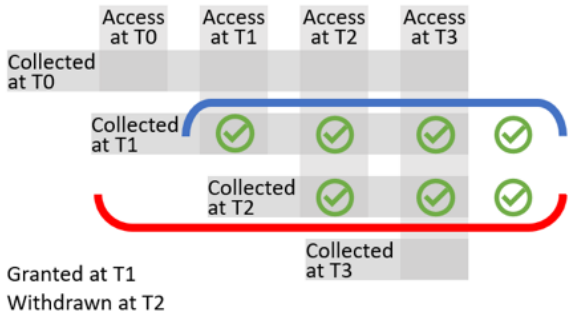

Figure 5. Consent non-retroactively granted, non-retroactively withdrawn

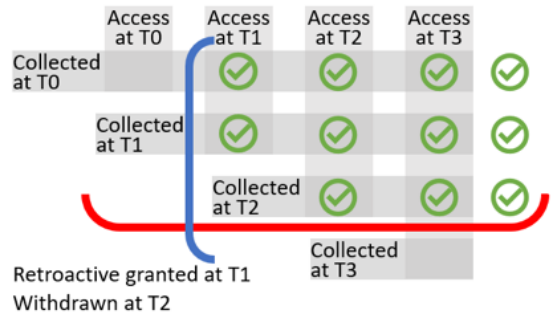

Figure 6. Consent retroactively granted, non-retroactively withdrawn

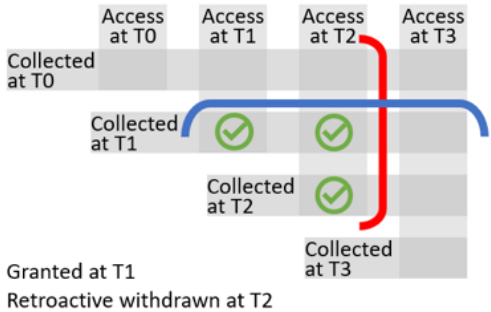

Figure 7. Consent non-retroactively granted, retroactively withdrawn

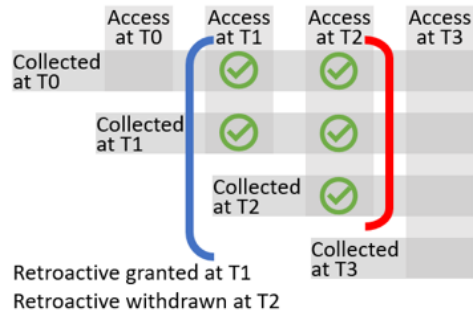

Figure 8. Consent retroactively granted, retroactively withdrawn

The difference between non-retroactively granted consent and retroactive consent is visible in Figures 4 and 5, respectively. In consent non-retroactively granted at time $T_1$, access is authorized to data collected at any future time $T_y$ for y >1, but not to data collected before $T_1$. In Figure 6, however, access to data collected before $T_1$ is authorized when retroactively granted.

Consent that is non-retroactively withdrawn and retroactively withdrawn is shown in Figures 4 and 6, respectively. In Figure 5, non-retroactive withdrawal at time $T_2$ means that access to data collected over the interval $T_1 \sqcap \neg T_3$ can continue at any future time $T_y$ for y >1, but collection cannot continue at any future time $T_z$ for z >2. In Figure 7, retroactive withdrawal limits access in the future, such that access is not authorized for any future time $T_z$ for z >2.

Figure 8 shows the authorizations for retroactively granted and withdrawn consent.

The European General Data Protection Regulation (GDPR) requires organizations to implement non-retroactive consent with non-retroactive withdrawal as shown in Figure 5. Notably, GDPR does not prohibit more restrictive consent, such as that shown in Figure 7, and GDPR is silent on whether organizations may request more relaxed access to historical data through retroactive consent, as shown in Figures 5 and 7.

## 5. Framework Application and Use Cases

In this section, we present five use cases that illustrate several types of evolution that can arise and how the framework can be used to evolve and verify consent under those use cases. In addition, we use a consent language syntax (see Appendix A) to concisely present events and time steps using the framework. In Appendix A, we present the denotational semantics that maps the language syntax into Description Logic defined by the formal framework. The language is supported by a tool, which is available online.[2] The tool can be used to check whether collection and access are authorized by consent, and to perform runtime monitoring that a collection and access log conforms to a current consent history. Logs can be translated into the scripting language for verification, or the framework's application programmer interface can be triggered by collection and access events to monitor consent at run-time (see Figure 2 for an example illustrating how the framework can be integrated into an architecture for run-time monitoring). The scenarios can be easily replicated using the off-the-shelf scenario scripts as described in the instructions provided within the repository.

We envision the following five use cases that policy authors may encounter when evolving their data practices. We identified these use cases when analyzing the consent life cycle presented in Figure 3, specifically, when considering the consequences of an organization evolving their data practices over time by adding and replacing services. These evolutions are triggered by changes in services or how data is managed, and they should not interfere with existing consent histories, i.e., they should not yield authorized actions without prior consent, nor yield unauthorized actions where prior consent was granted. In addition, users should be able to grant and revoke consent under both old and new policies.

- **Overlapping Authorizations** – organizations that collect data through multiple services, each covered by separate consents, can encounter overlapping authorizations. A user who revokes consent from one service could disrupt another service where consent remains granted, if the system does not distinguish overlapping consent and limit revocation, accordingly.

- **Refining Existing Data Type** – organizations will introduce new services and features that require new data types. A policy author may refine existing types by either adding new types to be included within existing types, or moving the existing types under new types. Changing the data type hierarchy must not introduce changes in the scope of previously granted consent.

- **Compartmentalizing Legacy Data** – organizations that migrate from non-granular policies to granular policies will need to preserve consent histories over data collected under legacy, non-granular policies. When completing this migration, the scope of previously granted consent must not change.

- **Classifying Data Types under Multiple Classes** – over time, organizations can classify existing data under additional classes to restructure their data management practices (e.g., introducing a financial data class to add greater security to protect this class). New policies can refer to these additional classes to simplify policy language.

- **Supporting New Purposes** – organizations may add new purposes for which data is used, but they must ensure that, only if a new purpose is similar to an original purpose, can a prior consent authorize collection and access under the new purpose, otherwise a new consent is needed for the new purpose.

---

[2] https://github.com/cmu-relab/consentsim.git

While the above five use cases were motived by business, legal and technological needs of the organization, they do not describe a complete enumeration of all possible use cases. We now review each of these use cases with specific illustrating examples using the framework.

## 5.1 Overlapping Authorizations

In the consent framework, data types and data purposes are hierarchical and thus it is possible for large organizations to collect data from a data subject under multiple, separate authorizations. When a data subject withdraws one consent, then questions arise: (1) can the data type that was previously authorized by the withdrawn consent still be collected under the remaining consent; and (2) can previously collected data under the withdrawn consent still be accessed under the remaining consent?

For example, consider a scenario wherein an organization requests consent to collect real-time location data for advertising purposes, which entails collecting precise location data every few minutes using a mobile device location service. The organization uses this data to compute the user's driving and walking routes by analyzing the device's real-time location and accelerometer data to infer that the user is driving or walking, and by inferring the route taken during a specific time interval. This scenario is expressed in the consent language as follows:

```
1   new data RealTimeLocation Data
2   new data DrivingRoute RealTimeLocation
3   new data WalkingRoute RealTimeLocation
4   new disjoin DrivingRoute WalkingRoute
5   new recipient Advertiser
6   grant RealTimeLocation datasubject1 Advertiser :consent1
7   collect DrivingRoute datasubject1 Advertiser
8   step
9   grant DrivingRoute datasubject1 Advertiser :consent2
10  collect DrivingRoute datasubject1 Advertiser
11  step
12  withdraw :consent1
13  assume false collect WalkingRoute datasubject1 Advertiser
14  assume true collect DrivingRoute datasubject1 Advertiser
15  assume true access DrivingRoute datasubject1 Advertiser T1
16  step
17  collect DrivingRoute datasubject1 Advertiser
18  step
19  withdraw retro :consent2
20  assume false collect DrivingRoute datasubject1 Advertiser
21  assume false access DrivingRoute datasubject1 Advertiser T4 T5
22  assume true access DrivingRoute datasubject1 Advertiser T1
```

In the scenario above, the default data type hierarchy is refined to include real time location and both driving and walking routes (lines 1-4), and a new recipient advertiser is added (line 5). After receiving consent from datasubject1 in time $T_1$ (line 6), the organization collects driving route (line 7), and later requests a separate, specific consent for the driving route at time $T_2$, at which point, they collect driving route a second time (lines 9-10). At time $T_3$, the data subject non-retroactively withdraws the first consent to collect real time location (line 12), leaving the second consent intact. At this time, the organization queries whether it can collect walking route, which is no longer authorized and thus assumed to be false (line 13), and driving route (line 14), which remains authorized under the second consent. Because the withdrawal was non-retroactive, the organization can continue to access the driving route that was collected in time $T_1$ under `consent1`, however, since `consent2` was withdrawn and `consent1` does not cover the time period of $T_3$ forward, the driving route collected during $T_4$ is no longer accessible as `consent2` was withdrawn retroactively.

In the consent framework, the data user only needs one authorization to guarantee collection and access. Thus, when consents are overlapping, the act to collect or access may be attributed to any of the overlapping consents that authorize the event.

## 5.2 Refining Existing Data Types

In the framework, consent, collection and access events are added to the knowledge base over time, which comprises the consent history and data collection and access log. When organizations release a new feature or product, however, they may need to update their policy by introducing new data types and purposes. If the new data types and purposes are subclasses of, and disjoint from, existing classes, then these additions are non-interfering, because they divide a previously undivided area in the domain of interpretation. However, if the new data types and purposes are superclasses of existing classes, then the change has the potential to interfere with the consistency of the consent history.

Consider, for example, an organization who previously collected `Location` data that was limited to location collected by a mobile device using cellular-based location. Later, the organization plans to divide this class into `CellularLocation` and `BluetoothLocation` to account for a new feature based on low-energy Bluetooth beacons. Because the consent history is based on `Location`, which would later come to mean CellularLocation, the organization must equate the concept `Location` to `CellularLocation`, and thus introduce a new generic class that subsumes these two more specific classes.

Recall, that consent covers generic classes, whereas acts of collection and access cover specific classes. In addition, answers to queries are vague and include uncertainty due to the open world assumption. Negative answers (the concept "bottom") guarantee that no consent is available in the knowledge base to perform the queried action, while positive answers (not "bottom") guarantee that at least a part of the queried action is covered by the consent. The data user must formulate queries to be as specific as possible and that cannot be further refined, so that a positive answer guarantees that existing consent covers the complete interpretation of the queried, intended action.

For example, consider an initial scenario wherein an organization first requests consent to permit the collection of `Location` data, which at the time means data collected using the mobile device's cellular network.

```
1   new data Location Data
2   new recipient Advertiser
3   grant Location datasubject1 Advertiser :consent1
4   collect Location datasubject1 Advertiser
5   step
6   withdraw retro :consent1
7   step
```

In the initial scenario, above, the policy author refers to this data type as simply `Location` (line 1). The data subject later grants their consent (line 3), after which the data is collected (line 4), and consent is later withdrawn retroactively (line 6). Upon developing and preparing to release the new feature, the policy author updates their policy to include new location meanings by introducing new data types as follows.

```
8   new data LocationV2 Data
9   new data CellularLocation LocationV2
10  new data BluetoothLocation LocationV2
11  new disjoint CellularLocation BluetoothLocation
12  new equiv Location CellularLocation
13  grant retro BluetoothLocation datasubject1 Advertiser :consent2
14  assume false access CellularLocation datasubject1 Advertiser
15  assume true access BluetoothLocation datasubject1 Advertiser
```

In this policy extension (lines 8-15), the organization introduces a new generic location concept, called `LocationV2`, and creates two refinements `CellularLocation` and `BluetoothLocation` (lines 8-10).

Next, they defined these two refinements as disjoint (line 11), before equating the legacy concept `Location` to the new `CellularLocation` concept (line 12). From that moment on, location data collected using the cellular network is tagged as `CellularLocation`, while location data collected using Bluetooth wireless is tagged as `BluetoothLocation`. Furthermore, previously collected data still exists under the legacy tag `Location`, however, because Location is equivalent to `CellularLocation`, it is also subsumed by the new generic concept `LocationV2`. Any previously consents granted to collect and access `Location` would follow to `CellularLocation`, and any new consents to `LocationV2` or `BluetoothLocation` would be restricted to the interpretation of those new classes. For example, the data subject grants consent on line (13), and the query for access to cellular location (line 14) is not guaranteed, because the interpretation of the prior withdrawal of consent to `Location` carries forward under the new policy. In contrast, the query to access `BluetoothLocation` (line 15) is covered by the new consent granted on line 13.

In the above example, the legacy concept `Location` describes a class of data that could be renamed to a single subclass. In the situation where previously collected location data, covers multiple, previously indistinguishable classes and the policy authors now wish to distinguish these types, they must instead preserve the consent log history by introducing a legacy class, which we now discuss.

### 5.3    Compartmentalizing Legacy Data

As systems evolve, under-specification in the system description and corresponding policy can introduce ambiguity and vagueness that is difficult to reconcile. This can arise due to operating a NoSQL database, for example, wherein multiple services use generic data tags to describe incoming data and the provenance needed to link data back to specific services cannot be reconstructed from logs. In this situation, the organization cannot distinguish the data stored in the database and thus they must create a legacy data class that subsumes all interpretations of this indistinguishable data so as to preserve the integrity of the consent history. For example, consider a new scenario in which an organization is collecting `TechnicalData`, which is otherwise indistinguishable, but which the organization chooses to distinguish going forward.

```
1   new data TechnicalData Data
2   new recipient Advertiser
3   grant TechnicalData datasubject1 Advertiser :consent1
4   collect TechnicalData datasubject1 Advertiser
5   step
6   withdraw :consent1
7   step
```

The scenario above in lines 1-7 describes a situation wherein the policy author begins with consent to collect a vague class `TechnicalData`, which is shared with the recipient Advertiser. Going forward, the policy author aims to distinguish the classes of data previously collected under this generic type, as described below.

```
8    new data PersonalInformation Data
9    new data NonPersonalInformation Data
10   new disjoint PersonalInformation NonPersonalInformation TechnicalData
11   grant NonPersonalInformation datasubject1 Advertiser :consent2
12   assume false collect TechnicalData datasubject1 Advertiser
13   assume true access TechnicalData datasubject1 Advertiser T1 T2
14   assume true collect NonPersonalInformation datasubejct1 Advertiser
```

The author begins by evolving the policy to distinguish between the disjoint types `PersonalInformation` and `NonPersonalInformation`, with `TechnicalData` becoming compartmentalized and retired as a data tag for collection events. Under the new policy, the data user can no longer collect the legacy data class `TechnicalData`, but they can continue to access previously collected data under time steps $T_1$ and $T_2$, but not under $T_3$ and beyond.

## 5.4 Classifying Data Types under Multiple Classes

Over time, organizations may wish to classify existing data using new attributes or qualities of the data, such as whether the data is personal, financial, identifiable, or anonymous, to name a few attributes. In an earlier example, we introduced `CellularLocation` and `BluetoothLocation`, which describe the device-based mechanisms by which location data is collected. As privacy preferences evolve, the organization may be interested in further distinguishing precise and coarse or imprecise location, noting the former introduces greater privacy risks and should receive increased restrictions on sharing. In the consent framework, organizations can redefine existing data types by updating the ontology in a monotonic way. This includes subclassing existing concepts under a new concept, without compromising the integrity of the knowledge base or invalidating past interpretations.

Consider an initial scenario in which only a device-based location classification exists, such that *BluetoothLocation* ⊑ *DeviceLocation*, and the policy author aims to introduce a refinement to the precision of the location-based data, wherein *BluetoothLocation* ⊑ *CoarseLocation* by a monotonic update of the knowledge base. This scenario is captured, below, where the old policy, including consent, collection and withdrawal, appears in lines 1-7, and the new refinement and new consent are introduced in lines 9-11.

```
1   new data Location Data
2   new data DeviceLocation Location
3   new data BluetoothLocation DeviceLocation
4   grant DeviceLocation datasubject1 Advertiser :consent1
5   collect BluetoothLocation datasubject1 Advertiser
6   step
7   withdraw retro :consent1
8   step
9   new data CoarseLocation Location
10  new data BluetoothLocation CoarseLocation
11  grant retro CoarseLocation datasubject1 Advertiser :consent2
12  assume true access BluetoothLocation datasubject1 Advertiser T1 T3
13  assume true access BluetoothLocation datasubject1 Advertiser T3
```

In the scenario above, access to `BluetoothLocation` is authorized both for previously collected data (line 5) and for newly collected data after the new policy takes effect (line 11). At the time of collection in $T_1$, the data subject was unaware that `BluetoothLocation` would be interpreted as `CoarseLocation`. Under the new policy, however, this interpretation is made explicit when the data subject grants consent later.

Under this policy change, new and old queries continue to work correctly, and new and old data will be classified consistently. However, a rollback of such changes implies a non-monotonic update and therefore counter actions as presented in Section 5.3 should be considered.

## 5.5 Supporting New Purposes

Over time, organizations may wish to evolve their policies by introducing new purposes. In 2012, the term Unique Device Identifier (UDID) appeared for the first time in the Waze policy in the following sentence: "Personally identifiable information may also be shared with Waze's other partners and service providers, with the express provision that their use of such information must comply with this policy. For example, Waze may share your mobile device's Unique Device Identifier (UDID) with ad networks, for the purposes described above." By including UDID in their policy, Waze may assume under US law that they have consent to share the UDID collected by their app with third party advertisement organizations for the purpose of targeted advertising.

The GDPR defines the purpose limitation principle, which means that data can only be used for a new purpose, (1) if the new purpose is comparable to the original purpose, (2) if consent is obtained for the new purpose, or (3) if there is a lawful basis. Herein, we formalize the first two conditions needed to guarantee the purpose limitation principle.

In the first situation, a consent already exists for a given purpose, which is defined by the conjunction of a data type and a recipient, $Purpose\_existing$: $D1 \sqcap R1$. In this situation, the purpose limitation principle allows the use of an existing consent, under the condition that the new purpose is included in the already consented purpose, formally $Purpose\_new \sqsubseteq Purpose\_existing$.

In the second situation, a new consent is required to cover the new purpose $Purpose\_new$: $D_x \sqcap R_y$. Formally $D_x \sqcap R_y \sqcap \forall authorizedBy.(ConsentSet)$ must not be an empty set, where $ConsentSet$ includes exactly and nothing more than all available consents.

For example, consider data initially collected and used for some specific purpose $dataCollection1 \in R1 \sqcap T1$, authorized by consent $T1 \sqsubseteq R1$. Later, the same data is accessed for a different purpose, $dataAccess1 \in R2 \sqcap T2 \sqcap access(dataCollection1)$. This operation is authorized if either: (i) the new purpose is comparable to (subsumed by) the original purpose, $R2 \sqsubseteq R1$, where $T2 \sqsubseteq T1$; or (ii) the new consent includes the new purpose $T2 \sqsubseteq T1 \sqcup R2$.

## 6. Scalability Evaluation

The consent framework and tool that supports the consent language syntax (see Appendix A) were used in a series of simulations to study the scalability of consent verification over time for a single data subject. This study was limited to the first two use cases for overlapping consent (see Section 6.1) and refining existing data types (see Section 6.2), since the subsequent three use cases are variations on the types of expressions covered by these use cases (e.g., adding new purposes is a refinement of the purpose hierarchy, similar to how adding new data types is a refinement of the data type hierarchy in Description Logic).

*Assumptions.* The study was designed with several assumptions in mind: (1) the study is limited to one application; (2) every day, the application collects and accesses data from the application user for which consent has been received; (3) each week, the organization updates their software to fix defects and add features, which could yield new data collections or accesses, and new recipients (e.g., by changing where data is hosted remotely, or changing advertisers); (4) every 90 days, the policy is updated, which requires a user to grant or withdraw consent; and (5) verification of data collection and access are performed once every day, typically at the start or end of the day when "housekeeping" functions are performed. The third assumption is based on how modern software engineering practices employ continuous integration and deployment. The fourth assumption is based on the analysis of the Waze privacy policy evolution, which shows that one policy can be updated as often as 3-4 times per year.

*Execution.* The simulations were conducted using an Acer TravelMate P238 Series laptop with an Intel i5 6200u 2.3 GHz processor, 8G RAM, and the Windows 10 Pro v2004 Build 19041.1052 operating system. The simulation was designed in Python v3.7.8 using OwlReady 2 v0.3, which was configured to use the default HermiT reasoner v1.3.8. The investigator closed all non-relevant applications while running the simulation, and each simulation was run five times and averaged the results for each simulation.

*Implementation.* Figure 9 shows three simple simulations conducted to measure the time to update the knowledge base, without synchronizing the base. Synchronization instructs the reasoner to build the complete entailment, which is needed to check whether a collection or access concept has an interpretation.

- **Steps**: in this simulation, each time step corresponds to a single day, in which only refinement of time is considered.

- **Nested Data Types**: each time step corresponds to a single day, in which a new data type is added to the most specific data type in knowledge base. In this simulation, no checks for collection or access are performed, and no consent events are recorded. This simulation demonstrates the effect of data type refinement on the time to build the knowledge base.

- **Nested Data Types and Recipients**: this simulation extends the previous one by also adding a new recipient every day, in addition to the new data.

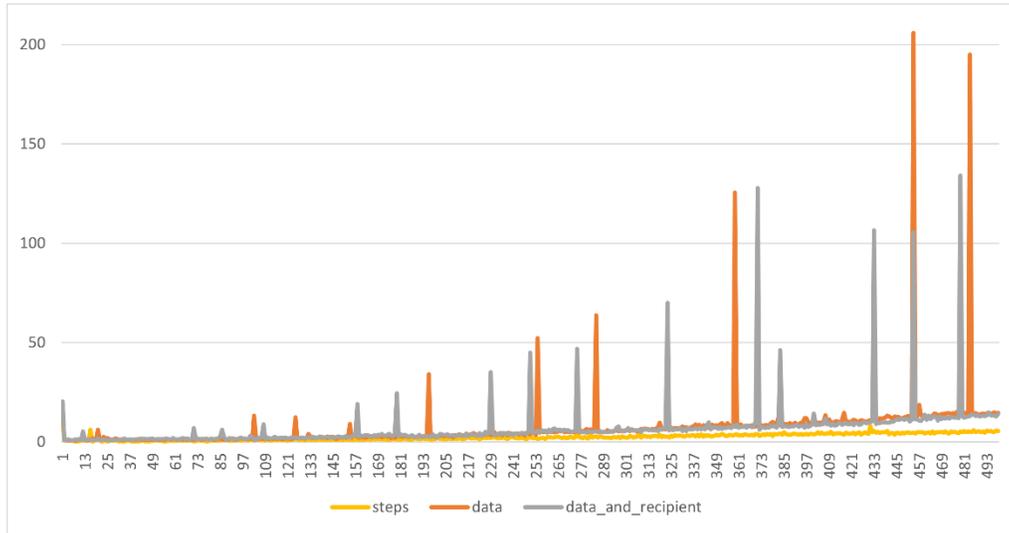

**Figure 9: Simulation results presenting the time to complete each time step in milliseconds**

In Figure 10, we present the results of five simulations in which the knowledge base is synchronized and queried to check for the possibility of collecting or accessing data:

- **Querying for Collection**: each time step corresponds to a single day, in which the knowledge base is queried to check for the possibility, given by the existence of a consent currently granted, to collect a specific data.

- **Refining Data and Querying for Collection**: each time step corresponds to a single day, in which a new data type is added to the knowledge base and a query is performed to check for the possibility, given by the existence of a consent currently granted, to collect a specific data.

- **Querying for Access**: each time step corresponds to a single day, in which the knowledge base is queried to check for the existence of a consent from the past or currently granted that allows to access data collected at a specific time in the past.

- **Refining Data and Querying for Access**: each time step corresponds to a single day, in which a new data type is added to the knowledge base and a query is performed to check for the existence of a consent from the past or currently granted that allows to access data collected at a specific time in the past.

- **Realistic:** once per day (1 time step), verification of a data collection and access is performed, then the application collects and accesses data from the application user; each week (7 time steps), the data ontology is refined by nesting a new data type to the most specific data type, and new recipients are added (e.g., by changing where data is hosted remotely, or changing advertisers); every 90 days (90 time steps), a new consent is granted and previous one is withdrawn, simulating a policy update and how data subjects react to the update.

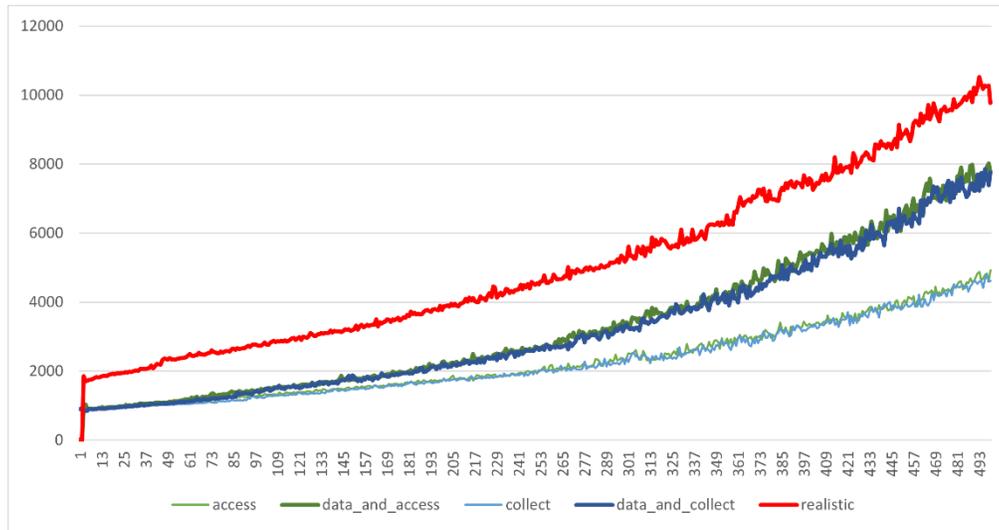

**Figure 10: Simulation results presenting the time to complete each time step in milliseconds**

**Interpretation of the results**

*Consideration.* As Figure 9 presents, the time to update the knowledge base is less than 1/50$^{th}$ of a second, apart from extemporary spikes probably caused by a systematic increment of the knowledge base. In contrast, synchronization, which is required to query the knowledge base, requires additional computation time. In Figure 10, the time to query the knowledge base for collection and access based on the realistic scenario at 90 days is 2.7 seconds for one data subject, and after one year is 6.7 seconds. Unlike access control, which guarantees that each data collection and access request is permitted at the time of the event, we assume that consent is checked once daily, such as at the start or end of the day. Moreover, we assume that checking collection and access are independent events from granting and withdrawing consent. Under the assumption of independence and checking consent once daily, organizations have 24 hours to process a consent request, before collection and access would be prohibited. In practice, however, only three types of events – a policy evolution or a data subject granting or withdrawing consent – would change the authorization of collection or access to data. Thus, organizations could assume the events are dependent, by binding these three events to a new check (e.g., instead of checking daily, the check may be performed weekly or every few months, when software or policies are updated.) We envision a large design space in which organizations may or may not have the ability to bind such events, and therefore they may need to process events in batches, increasing the time to process a consent withdrawal. To our knowledge, the GDPR does not restrict the time to process a withdrawal; however, regulators typically do allow organizations a "reasonable" time to process requests.

*Complexity.* A number of factors can affect the complexity of reasoning under the consent framework, including: the number of data subjects whose information is being processed; the pace of evolution in business practices, such as the number of new data types being processed; and the number of third-parties participating in processing. For some organizations, parallel processing and few changes to business practices and suppliers would result in less time to check collection and access. Larger organizations generally have more infrastructure to support compliance. Regardless, organizations could implement optimizations to reduce complexity. In addition to making collection and access checks dependent on policy and consent changes, organizations could limit the granularity of the data types and purposes being processed. This generality would yield less evolution, fewer policy notifications to data subjects, and thus fewer checks on collection and access. In addition, organizations could use retroactivity to reduce the complexity of reasoning. Non-retroactive policies can create discontinuity in data accessibility, which requires additional checks into the future. Retroactive policies can consolidate authorizations by either making previously collected data uniformly accessible, or uniformly inaccessible. Retroactive policies that increase access, however, may have a negative impact on privacy and thus may lead to a larger number of withdrawals by data subjects. Thus companies, must balance privacy protections and vulnerabilities that arise from retroactive policies that increase accessibility.

In summary, the simulations show the feasibility of checking collection and access periodically at runtime using the consent framework, given a realistic scenario for evolving policies over time. Depending on variables with the organization, including how their applications are designed, optimizations may be needed to scale the consent framework to a specific organizational setting.

## 7. Threats to Validity

*Construct validity* addresses whether what we measure is actually the construct of interest [53]. In empirical research, this concerns the design of surveys and experiments. In formal modeling, this may refer to the realism of the primitives and their alignment with real world phenomena. The framework introduces concepts such as user acts to grant and withdrawal consent, organizational collection and access to data. Prior studies on privacy policy language identified the actions of collection, use and sharing as the top three most prominent policy descriptions [18, 19]. The "use" action is used to link data type to data purpose in policies. In the consent framework, we distinguish collection and access, wherein access combines uses and sharing without distinction. If an organization needs to distinguish data sharing or transfers to third parties, then access could be sub-classed to distinguish two disjoint forms of access: internal access, and third-party sharing.

While granting and withdrawing consent are recognizable in user interfaces (e.g., click-through agreements, opt-out checkboxes), application functions for data collection and access are numerous and often distributed across complex service compositions. For example, web applications include collecting form data during online registration, and telemetry frameworks, some of which are third-party services, that collect a user's IP address, referrer host, and which HTML elements a user "mouses" over, among other data types. Mobile applications further have access to mobile device sensors, and can read and write data to device or network storage. The framework concepts of collection and access, when combined with data type concepts, are intentionally coarse-grained to cover such actions without the burden of enumeration. The evaluation use cases that illustrate evolution, however, demonstrate the framework's resilience to increasing granularity when adding new data types and purposes.

*Internal validity* concerns whether effects properly follow from the causes in an empirical study [53]. In formal modeling, this concerns the correctness of the model and model inferencing. To mitigate threats to correctness, we limit our framework to concept satisfiability and subsumption in the ALC family of Description Logics, which has been proven correct through tableau expansions [10]. The consent framework then introduces axioms in Section 4 to model concepts for consent granting and revocation, data collection and access, and time to yield consent histories over a monotonically increasing TBox and ABox. To test framework, we identified and examined five policy evolution use cases to show that each evolution did not yield changes in the TBox and ABox interpretation given prior consent histories (i.e., no change allowed a previously unauthorized collection or access, and no change prohibited a previously authorized collection or access.)

*External validity* refers to the extent to which we can generalize the results [53]. The generalizability of the framework was evaluated in two ways: (1) we studied the framework under five policy evolution use cases; and (2) we conducted a simulation to evaluate the scalability of the approach. With regard to (1), the use cases were selected based on the authors familiarity with the domain, recognizing that the problem of consent evolution is still quite new. For example, among the six consent scenarios shown in Figure 4, only two are required by the GDPR and actively in use to our knowledge (non-retroactive granting, and non-retroactive withdrawal). Thus, we expect the use cases to be incomplete with regard to the possible use cases that organizations may encounter in the future. With regard to (2), the simulation examines the scalability of the theorem prover to reason over policy evolutions of varying complexity. However, different factors could affect the interpretation of these results, based on how an organization would practically adopt the framework for monitoring consent at runtime. For example, collection and access queries could be batched together and triggered only when consent is changed on a per-user basis and used to check and update classical role-based access control rules. In addition, the assumptions in the realistic scenario, which include the frequency of policy evolution, data collection and access, may not match the situation faced by other companies. Therefore, the scalability results could be different under different assumptions.

## 8. Related Work

In this section, we discuss foundational and related work.

### 8.1 Temporality in Description Logic

Description Logic (DL) are a subset of first order logic, intended as a general-purpose language for knowledge representation, where decidability is valued over expressiveness. The components of description logic are: (i) concepts, (ii) their relations or properties, and (iii) individuals. When using DL to represent an application domain, definitions of concepts and properties compose the TBox, while assertions about individuals and their concepts and properties compose the ABox.

Temporal representation is not directly supported by DL. However, time and temporal concepts can be modeled. The OWL-Time ontology [28] provides concepts related to time representation, however it does not specify how to use these concepts, nor how to reason over such concepts. In general, temporal representation focuses on instants and/or intervals. In a point-based representation, relations between instants are "before", "after", and "equals." In an interval-based representation, relations can get up to the 13 pairwise disjoint Allen's relations [3] showed in Figure 9.

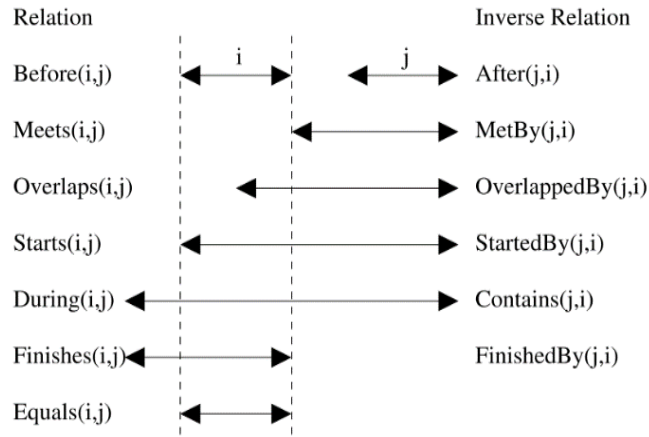

**Figure 9: Allen's temporal interval relations**

In the case of numerical representation of time, Allen's relations can be easily inferred. For qualitative representation, by assertion of Allen's relations, inferring non-declared relations or checking consistency is an NP-hard problem.

Apart from missing representation of time, DL formalisms also miss constructs to represent the evolution of concepts and their properties in time. Because DL only supports binary relations, temporality is not easily encoded as a dimension of an existing relation. For example, a data collection can be expressed as a relation between a data type and a user, however, to include also the time of collection we would need a ternary relation.

Many approaches have been proposed to address such problems [7]. Versioning has been discussed in [29], which proposes to create a copy of the ontology at every change. The n-ary relations approach [34] and 4D-fluents [52] are two alternative approaches to represent evolution of concepts. N-ary suggests representing a temporal ternary relation (object, verb, predicate, time) as a concept itself representing the verb, with properties to relate it to the object, the predicate, and the time. The 4D-fluents approach represents a temporal relation as a 4-dimensional object, which includes time-specific temporal versions of the object and the predicate, where the original relation is now expressed between the temporal versions of the concepts. With respect to the n-ary approach, the 4D-fluents approach suffers from proliferation of objects (two additional objects for each temporal relation). The n-ary approach, on the other hand, suffers from data redundancy in the case of inverse and symmetric properties (e.g., the inverse of a relation is added explicitly twice).

**Relation to Temporal Logic**

In the representation of consent evolution, collection defines a time instant and consent defines an interval between consent approval and withdrawal. Our approach is based on the representation of non-overlapping time intervals, where data are collected within one of these intervals, while a sequence of intervals defines a consent.

## 8.2 Privacy preferences and user personalization

Recently, organizations have pursued advanced personalization of user experiences to strengthen their relationship with users [4, 10, 46]. Personalization is commonly intended as customized or customizable user experiences, based on user's behavior or preferences. With respect to privacy preferences, users can explicitly request how or when their personal data will be used. For example, users may want to exclude their browser search history from the dataset used by websites to show targeted advertisements, or they may want to restrict access by other users to specific subsets of their personal information.

Ackerman at al. in [1] present survey results of user attitudes toward various E-commerce privacy concerns. Information about user attitudes can inform how designers choose which information to regulate using preferences. The authors found that user attitudes vary around current information practices, and thus they propose privacy clusters for individuals, labeled fundamentalists, pragmatics, and unconcerned. Some preferences are shared by all respondents, such as automatic transfer and unsolicited communications are disliked, while persistent identifiers are sometimes tolerated. Much of the discomfort with the Web today results from not knowing, or not trusting the information practices of a site [1].

Teltzrow et al. in [47] categorize personalization systems according to the data that they collect and then review surveys on privacy concerns, showing that users' privacy concerns have a direct impact on personalization systems. Two approaches are discussed to implement personalization while alleviating the concerns of users: one based on policies and laws, and the other based on anonymity.

Spiekermann et al. in [45] describe the conflict between the general desire of people to protect their privacy and an organization's aim to improve customer retention using service personalization. The contribution consists of an empirical study that measures the actual behavior of people versus their stated privacy preferences. The authors categorize approaches to address the privacy issue in three categories: law (the EU model), self-regulation (the US model), and standards. A simple-to-use identity management system is suggested as an important privacy technology of the future as an alternative to the limitations of anonymity and private credentials.

The Platform for Privacy Preference (P3P) [23] is the reference platform for privacy preferences on the web. With P3P, web browsers and web servers can communicate a user's privacy preference and compare those preferences with a website's privacy policy. The P3P can be used to report misalignments, however, the preferences are never used to modify a website's privacy settings as part of P3P. P3P is a form of electronic privacy policy or notice, and Schaub et al. concluded a study exploring the broader design space of privacy notices [42].

Despite the availability of privacy preference platforms, Acquisti et al. in [2] show that people cannot always be counted on to set their own privacy preferences in a self-interested fashion. For the authors the reasons are three-fold: (i) individuals are uncertain about their privacy behaviors and the consequences of those behaviors; (ii) the context-dependency of individual concerns about privacy; (iii) and the degree to which privacy concerns are malleable by commercial interests. Acquisti et al. conclude that policy approaches that rely exclusively on informing or "empowering" the individual are unlikely to provide adequate privacy protection [2].

In the mobile app world, where something like a privacy platform for preferences is the permissions system offered by the operating system. Lin et al. [30] investigate the feasibility of identifying a small set of privacy profiles as a way of helping users manage their mobile app privacy preferences.

Privacy risk concerns the likelihood that an individual's personal information could be used in a harmful manner, and the impact of that harm[3]. Daniel Solove analyzed U.S. court decisions concerning personal privacy and proposed a taxonomy of privacy harms, that include surveillance, forced disclosure and blackmail [44]. Bhatia and Breaux designed and evaluated a framework to measure the risk to personal privacy of disclosing different types of information [14]. Because privacy harm is difficult to measure, they designed a scale that can be used to increase or decrease the perception of harm based on social and physical proximity to the harm. They further propose that developers can choose preferences to enhance personalization by measuring privacy risks across their userbase, and then choosing which information types can be restricted based on variable perceptions of risk.

Studies have been done in analyzing the effects of the GDPR among web privacy policies around the globe, specifically with respect to preferences for cookies and user tracking. Zaeem et al. in [54] quantified these effects by comparing policies with the use of PrivacyChecker, a data mining tool to process textual privacy policies. Degeling et al. in [25] analyzed changes to websites privacy policies at the time when the GDPR came into effect and their approach to consent management for cookies. They argue against the confusion that still exists after the GDPR about how opt-ing out from third party cookies should be enforced on the web. Nouwens et al. in [33] present a study about the effects of privacy pop-up on the behavior of web users, showing that small modification to the graphical interface of a website can largely affect behavior of users with respect to privacy.

Pöhls in [37] consider the problem that third-parties data processing companies may have in dynamically verifying and proving continuously changing privacy preferences. Since preferences are not directly managed by third-parties it may be impossible for them to provide a proof of consent, especially considering that preferences are often modified by the users through the first-party company that is collecting the data, which may omit to notify the third-party about changes in the preferences. In the solution proposed by the authors, consent and privacy preferences are always associated within the data, which is structured as a tree. The data structure is then hashed and digitally signed with a trusted public key certificate.

Appenzeller at al. in [5] propose an architecture for the management of consent on health data for research purposes. The solution gives to the patient direct control on his data and consents.

**Relation to Privacy Preferences**

We described above a limited selection of works in privacy preferences, most of which are about measuring how people feel and perceive privacy with respect to their preferences. Together with personalization, these two themes are very important in the research community.

Our work on consent described herein complements this prior work on privacy attitudes, preferences or risk, since it provides organizations a robust mechanism for managing changing consent options and user preferences over time. For example, privacy attitudes and risk may be subject to recent events, and to discovering the benefits and harms of how systems collect and use personal information [1, 11]. People do react to opportunities to change consent under different settings, and privacy preferences can give organizations more flexibility (e.g., versus withdrawing consent and leaving the service, a user could modify a single preference that only degrades the service quality for that individual user). Our framework is a first result to help organizations understand their opportunities to maintain an appropriate level of access to user data while increasing their user-base and personalization-quality.

---
[3] https://csrc.nist.gov/publications/detail/sp/800-37/rev-2/final

## 8.3 Formalization of privacy policy

Research to formalize privacy policy has sought to encode important design principles or to answer basic questions. For example, contextual integrity states that information flows conform to norms consisting of a sender, a receiver and who information is about, and was formalized in temporal logic to detect inconsistencies between positive and negative norms [12]. Checking for policy conflicts, however, requires abstract or vague expressions of data types and purposes, and which is supported by description logic [18]. Moreover, description logic can be used to formally trace data flows across third-party policies, and to verify the purpose limitation and data minimization principles, also called the use and collection limitation principles [13]. J. Bhatia et al. in [15], addressing the need for collection limitation, presented a case study about data purposes in privacy policies, in which they identified six exclusive purposes found in the policies. Then in [16] they analyze the semantic of public privacy policies and report about incompleteness of data practices description in the policies, arguing about possible erroneous comprehension by the data subject.

Vanezi et al. in [51] proposes the use of Privacy-Calculus in a model-driven development approach to statically check the compliance with respect to the GDPR, and more specifically, on lawfulness of processing based on consent, consent withdrawal, and right to erasure.

**Consent monitoring**

Artfelt et al. in [6] analyze GDPR articles and propose a formalization of articles related to monitorable system actions. The proposed formalization is based on first-order temporal logic. They do not focus strictly on consent, but more in general on GDPR data processing principles. When they come to consent, however, even if they claim use and collection to be conflated into processing within the GDPR, they distinguish between them and only consider consent for use. Collection is considered separately and then related to consent with the data minimization principle, even though data collection before consent is still allowed. Additionally, granting and withdrawal are always retroactive events, in the sense they always apply to both new and already collected data. In this sense, our framework, which focus is solely on consent, and specifically on evolution, is more generalizable, allowing combinations of retroactivity and non-retroactivity in consent granting and revoking. Moreover, purpose is not considered into the formalization of consent, possibly misleading limitations on consent imposed by the GDPR.

In comparison, Bonatti at al. in [17], similarly to us, considered also purpose of data processing, in addition to data category, data subject, and recipient. Moreover, in their analysis of GDPR they also included cross-border data transfer and security measures in the context of personal data processing. They propose the SPECIAL language for consent (presented in BNF format) with automated compliance checking (based on OWL). Also in this case, policies by default refer to the use of data.

**Blockchain and smart contracts**

Blockchain and smart contracts have also been used to formalize and monitor the actions access, store and transfer over personal data and to encode answers to GDPR-related questions in an immutable contract [11]. In this prior work, however, the issue of granting and revoking consent was not addressed, including the effects of retroactivity. N. Truong et al. in [50] propose a blockchain-based solution to GDPR compliance, in which adhering companies use smart contracts to upload on the blockchain all data and consent activities, which are then automatically verified and certified on immutable records.

**UML Framework(s)**

Torre et al. in [49] propose an UML diagram to capture requirements for GDPR compliance as a first step toward an automated engineering method for companies. The authors identify future directions including a domain-specific language similar to OCL but more understandable by legal experts; goal modeling to reason on decomposition and delegation of tasks and responsibilities between companies; and Machine Learning (ML) and Natural Language Processing (NLP) to support automated processing of legal policies.

**Relation to Formalization**

In this paper, the consent framework addresses Article (5) in GDPR with an additional emphasis on policy evolution and how that can lead to non-compliance. With regard to prior work, Artfelt et al. in [6] proposes a comprehensive formalization of GDPR, including Article(5).

## 8.4 Access Control

Access control mechanisms are commonly adopted by system administrators to regulate access to resources, such as data or application access. Access control is generally configured to model corporate structure and personnel roles. When enforced on databases, access control mechanisms regulate only access to the data and not how the data is used.

Sandhu at al. [41, 40] proposed Role-Based Access Control (RBAC), which has become the industry standard for access control. In RBAC, authorizations are granted based on the role of the users. Organizations may statically assign roles to employee based on their job functions. One limitation of RBAC, however, is the impossibility to allocate authorizations dynamically, based on the activity for which data access is required. Task-Based Access Control (TBAC) was first proposed by Thomas and Sandhu [48] to satisfy the needs of enterprises in the era of agent-based distributed computing, by moving from the classical subject-oriented authorization model, toward task-oriented models.

As demand for richer access control models continued, Park et al. in [36] introduced the Usage Control (UCON) mechanism, to unify approaches for access control, trust management, and digital rights. Trust management and digital rights differ from access control, in that the former focuses on authentication of users in open environments, while latter focuses on unauthorized use of the data on the client system (a concept more similar to task-based access control). Later, Byun, Bertino, et al. in [21] proposed Purpose-Based Access Control (PBAC) to control data access based on intended purpose associated with the data, in which metadata about purposes is associated with a given data to specify its intended use. Finally, unlike data purpose, which assumes independent activities for which data is used, Nguyen, Park et al. in [35] proposed Provenance-based Access Control (PBAC), which is a mechanism that utilizes the provenance of data [20] to control the access to the data. Davari and Bertino in [24] propose a decentralized access control framework that supports GDPR requirements, based on XACML.

Other solutions exist for specific applications, such as Proximity-based Access Control [27] which has been proposed for emergency situations, wherein people are constantly moving and need to quickly access resources, which are located in different physical locations.

**Relation to Access Control**

The formal consent framework proposed herein is an abstract layer resting atop one or more of these technical solutions. The framework aggregates access control rules into a consent mechanism that is framed by the data controller through data purpose, controlled or actuated by the data subject through granting and withdrawing consent, and implemented or enforced by the data processor using one or more access control mechanisms. While the framework is mechanism-agnostic, the purposes expressed in the framework can be used with roles, tasks or purposes found in RBAC, TBAC or PBAC.

Unlike access control, wherein rules are written to allow or deny access, granting and withdrawing consent are modifications to separate, but dependent, authorizations to collect and access data over time. There is no mechanism to directly deny access to data in the consent framework, only a means to grant and limit authorizations. Granting consent yields new permissions to collect and access data, whereas withdrawing consent simply removes some or all of those permissions. This is similar to a deny-first, allow-later paradigm, in which the absence of authorization means access is unauthorized. Moreover, authorizations may be overlapping in the type of data, purpose of use, and time. Time, in particular, is shaped by retroactivity, which can be assigned to the granting and/or withdrawal of consent.

## 9. Conclusion

We presented a framework for the representation and analysis of evolving consent to support organizations in better understanding how consent changes affect their ability to access data in a legally compliant manner. This work is an attempt to show the complexity of consent evolution and demonstrate the capability and limitations of description logic in modeling consent evolution. Under the GDPR, the framework can be used to support organizations through the framework language as follows. As data controllers, organizations should document their data purposes using a classification system, which could be expressed in description logic using the language, and create an accounting in the language of points within their enterprise architecture where consent is collected, e.g., across web sites, mobile applications, etc. Changes to the consent text at these points should correspond to changes expressed in the framework language. As data processors, organizations can analyze database logs for collection and access events for runtime monitoring to detect violations. To avoid violations, however, organizations can require developers to use collection and access enforcement points. For example, developers and users of Application Programmer Interfaces (APIs) can document their APIs using code-level annotations that can be statically checked against expressions in the language indicting the developer's design intent (e.g., to collect wireless location for advertising purposes). Each "check" can dynamically report in real-time to the developer whether *any current consents* permit the intended collection or access, as well as the proportion of the userbase and type of users who have opted-in. With a framework implementation similar to the one described above, organizations can more precisely demonstrate their GDPR compliance to regulators, and regulators can check consent histories as evidence of compliance.

While the consent framework includes a formalization of consent, data collection and data access, and tools to automatically detect policy violations, there is a limitation that the framework users must instantiate the framework within their organization before it can be used. To perform design-time verification, this instantiation requires designing scenarios that reflect planned data practices, including adding new data types and roles to a new scenario script. To perform monitoring, this requires integrating the framework into the organizational architecture (e.g., see Figure 5), which may require writing drivers to parse log files to yield data collection and access events. In both cases, the framework implementation and tools provide a scriptable interface that organizations can use to reduce this manual effort. That said, privacy protection has a cost to the data subject and to the organization, that is not easy to eliminate. No solution is fully automated, and there is always a cost for integration into existing infrastructure.

Future work includes a further evaluation of the scalability of the approach and in particular how multiple consents can affect big data applications, such as machine learning. In organizations with a broad range of services, or in data brokers, data may be pooled under different authorities and data purposes. In this setting, the data user may be interested in examining cross-domain applications, such as how one dataset, such as purchase histories, reveals insights into a person's health using health data. The results of such analysis should be restricted to the intersection of data purposes assigned to the individual datasets. For example, whereas purchase histories may be treated as marketing data used for advertising purposes, and health data is used for diagnostic and treatment purposes, then an intersection may be empty, if these two purposes are disjoint. Alternatively, if the intersection is described by a purpose for advertising treatments options, then the results may be limited to that purpose. Other purposes, such as performing adjustments to health insurance premiums, however, would be excluded. We believe the consent framework can be used to model this kind of analysis and to restrict big data analysis to be GDPR compliant, while still permitting legal data exploration.

Finally, the relationship to privacy preferences is worth further consideration. Unlike consent, preferences can turn on or off data collection and access at a specific time, which provides users greater control over their privacy. However, what constitutes the meaning of "on" and "off" can also be affected by retroactivity in consent and withdrawal. By affording users greater control over their data, organizations can offer preferences to adjust access to especially sensitive data types. Similarly, organizations may want to separate sensitive data types from non-sensitive data types using separate user consents to avoid having users opt-out entirely from using a service. In addition, by offering retroactive withdrawal for sensitive data types, which is a stronger protection than what is required under GDPR, organizations can allow users to exclude sensitive data from future uses after leaving a service. Upon those same users returning in the future and agreeing to retroactive consent, they would regain access to the service with their prior data intact, which is not the case under the Right to Be Forgotten.

## Appendix A: Consent Framework Language

The consent framework language is used to express scenarios consisting of a sequence of data subject and data user actions. The formal semantics for the language are expressed in the Web Ontology Language (OWL) 2.0. The language is supported by a tool that builds the consent history in OWL from the scenario, which can then be used to automate queries over the history. The language abstract syntax is expressed below in the Extended Backus Naur Form (EBNF): words in angled brackets are production rules, lowercase words are keywords in the language, and words with capitalization are ground terms that correspond to values in data, data subjects and recipients.

```
<S>        := <new> | <rename> | <grant> | <withdraw> | <collect> | <access> |
              step | <assume>
<new>      := new data DataClass DataSuperClass | new recipient RecipientClass
              | new disjoint DataClass1 DataClass2 | new equiv DataClass1 DataClass2
<grant>    := grant (retro)? <datadesc> :ConsentIndiv
<withdraw> := withdraw (retro)? :ConsentIndiv
<collect>  := collect <datadesc>
<access>   := access <datadesc>
<datadesc> := DataClass DataSubjectIndiv RecipientClass
<assume>   := assume true <action> | assume false <action>
```

The abstract syntax is assigned a denotational semantics [43] based on a semantic domain expressed in Description Logic (DL). The semantic domain consists of DL classes, DL operators for union, intersection and negation, and a program state maintained by the language interpreter, which is a tuple (*KB*, *t*, *e*, *class*, *indiv*, *q*) for a DL knowledge base *KB*, the current time step *t*, the next event counter *e*, the *class* and *indiv* functions, and a query result *q*. As time advances, the time step *t* is incremented by 1, and for each new collection and access event, the even counter *e* is incremented by 1. To facilitate translating concept and individual names expressed in the language to classes and individuals in TBox and ABox, we define the functions class → Names × Classes and indiv → Names × Individuals, respectively. Finally, the language supports Boolean queries to test the authority to collect and access data, which results in changes to the Boolean value of *q*. The initial values for the program state are ($\{T_1 \sqsubseteq Time, T_2 \sqsubseteq T_1\}$, 1, 1, {(Time, Time), ($T_1$, $T_1$), ($T_2$, $T_2$), (Data, Data)}, {}, False). Finally, we use the & operator to express string concatenation, which we use when composing new class and individual names. The language syntax is mapped to semantic functions used to build the knowledge base. Given a program *P* expressed in the language syntax, the meaning of *P* is defined by evaluations of individual expressions *E*, including sequenced in *S*, through a series of *update* and *express* semantic functions as follows.

$meaning[\![P]\!] = KB$ where
$\quad evaluate[\![P]\!](\{T1 \sqsubseteq Time, T_2 \sqsubseteq T_1\}, 1, 1, \quad \{(Time, Time), (T_1, T_1), (Data, Data)\}, \{\}, False))$
$\qquad = (KB, t, e, class, indiv, q)$

$evaluate[\![E\ S]\!](KB, t, e, class, indiv, q) = update[\![E]\!] \circ evaluate[\![S]\!]$

$update[\![new\ data\ p_1\ p_2]\!](KB, t, e, class, indiv, q) = (KB \cup \{class(p_1) \sqsubseteq class(p_2), class(p_1)$
$\qquad \sqsubseteq Data, class(p_2) \sqsubseteq Data\}, t, e, class, indiv, q)$

$update[\![new\ recipient\ p_1\ p_2]\!](KB, t, e, class, indiv, q) = (KB \cup \{class(p_1) \sqsubseteq class(p_2), class(p_1)$
$\qquad \sqsubseteq Recipient, class(p_2) \sqsubseteq Recipient\}, t, e, class, indiv, q)$

$update[\![new\ disjoint\ p_1\ p_2]\!](KB, t, e, class, indiv, q) = (KB \cup \{class(p_1) \sqcap class(p_2) \to$
$\qquad \bot\}, t, e, class, indiv, q)$

$update[\![new\ equiv\ p_1\ p_2]\!](KB, t, e, class, indiv, q) = (KB \cup \{class(p_1)$
$\qquad \equiv class(p_2)\}, t, e, class, indiv, q)$

$update[\![grant\ D\ S\ R\ C]\!](KB, t, e, class, indiv, q)$
$\qquad = (KB \cup \{indiv(C))$
$\qquad \in express[\![auth\ D]\!](class, indiv) \sqcap express[\![open\ interval]\!](t)$
$\qquad \sqcap express[\![auth\ R\ S]\!](class)\}, t, e, class, indiv, q)$

$update[\![grant\ retro\ D\ S\ R\ C]\!](KB, t, e, class, indiv, q)$
$\qquad = (KB \cup \{indiv(C))$
$\qquad \in express[\![auth\ D]\!](class, indiv) \sqcap express[\![retro\ open\ interval]\!](t)$
$\qquad \sqcap express[\![auth\ R\ S]\!](class)\}, t, e, class, indiv, q)$

$update[\![withdraw\ C]\!](KB, t, e, class, indiv, q)$
$\qquad = (KB \cup \{indiv(C)) \in express[\![close\ interval]\!](t)\}, t, e, class, indiv, q)$

$update[\![withdraw\ retro\ C]\!](KB, t, e, class, indiv, q)$
$\qquad = (KB \cup \{indiv(C)) \in express[\![retro\ close\ interval]\!](t)\}, t, e, class, indiv, q)$

$update[\![collect\ D\ R\ S]\!](KB, t, e, class, indiv, q)$
$\qquad = (KB \cup \{express[\![collect\ D\ R\ S]\!](t, e, class, indiv)\}, t, e + 1, class, indiv$
$\qquad \cup \{("event"\&e, e)\}, q)$

$update[\![access\ D\ R\ S]\!](KB, t, e, class, indiv, q)$
$\qquad = (KB \cup \{express[\![access\ D\ R\ S]\!](t, e, class, indiv)\}, t, e + 1, class, indiv$
$\qquad \cup \{("event"\&e, )\}, q)$

$\quad update[\![assume\ T\ access\ D\ R\ S]\!](KB, t, e, class, indiv, q)$
$\qquad = (KB, t, e, class, indiv, \neg express[\![T]\!] \wedge KB \vDash express[\![access\ D\ R\ S]\!] \equiv \bot)$

$update[\![assume\ T\ collect\ D\ R\ S]\!](KB, t, e, class, indiv, q) = (KB, t, e, class, indiv, \neg express[\![T]\!] \wedge KB$
$\qquad \vDash express[\![collect\ D\ R\ S]\!] \equiv \bot)$

$update[\![step]\!](KB, t, e, class, indiv, q) = (KB \cup \{"T"\&(t + 2) \sqsubseteq "T"\&(t + 1)\}, t + 1, e, class, indiv, q)$

$express[\![auth\ D]\!](class, indiv) = Consent \sqcap \forall authorizes.(class(D))$

$express[\![open\ interval]\!](t)$
$\qquad = \forall authorizes.\begin{pmatrix} (Collection \sqcap \exists collectedAt.("T"\&t)) \\ \sqcup (Access \sqcap \exists accessedAt.T\&t \sqcap \exists collectedAt.("T"\&t)) \end{pmatrix}$

$express[\![retro\ open\ interval]\!](t)$
$\qquad = \forall authorizes.\begin{pmatrix} (Collection \sqcap \exists collectedAt.("T"\&t)) \\ \sqcup (Access \sqcap \exists accessedAt.("T"\&t) \sqcap \exists collectedAt.Time) \end{pmatrix}$

$express[\![close\ interval]\!](t) = \neg \forall authorizes.\begin{pmatrix} (Collection \sqcap \exists collectedAt.("T"\&t)) \\ \sqcup (Access \sqcap \exists collectedAt.("T"\&t)) \end{pmatrix}$

$express[\![retro\ close\ interval]\!](t)$
$\qquad = \neg \forall authorizes.\begin{pmatrix} (Collection \sqcap \exists collectedAt.("T"\&t)) \\ \sqcup (Access \sqcap \exists accessedAt.("T"\&t) \sqcap \exists collectedAt.("T"\&t)) \end{pmatrix}$

$express[\![auth\ R\ S]\!](class, indiv) = \forall authorizes.\exists collectedBy.(class(R)) \sqcap about.(indiv(S))$

$$express[\![access\ D\ R\ S]\!](t, e, class, indiv) = indiv("event"\&e)$$
$$\in Acceess \sqcap \exists accessedAt.("T"\&t \sqcap \neg"T"\&(t+1)) \sqcap \exists accessedBy.class(R)$$
$$\sqcap\ about.indiv(S)$$
$$express[\![collect\ D\ R\ S]\!](t, e, class, indiv) = indiv("event"\&e)$$
$$\in Collection \sqcap \exists collectedAt.\left("T"\&t \sqcap \neg"T"\&(t+1)\right) \sqcap \exists collectedBy.class(R)$$
$$\sqcap\ about.indiv(S)$$
$$express[\![T]\!] = true,\ if\ T = \text{"true,"}\ else\ false$$

An example of a simple scenario expressed in the language, below in lines 1-7, begins with creating a new data type `Location` as a subclass of `Data`, and a new recipient `Advertiser`. Next, in line 3, the data subject (`datasubject1`) grants non-retroactive consent to the `Advertiser` to collect and access Location. Non-retroactive consent and withdrawal are the default modalities. This consent is labeled `:consent1` to support a future withdrawal. In line 4, the scenario tests an assumption expecting it to be true that `Location` data can now be collected by the `Advertiser` from the `datasubject1`. In line 5, the data user collects this data type, and line 6, the time is advanced to the next step. Finally, in line 7, the data subject withdraws `:consent1` retroactively.

```
1  new data Location Data
2  new recipient Advertiser
3  grant Location datasubject1 Advertiser :consent1
4  assume true collect Location datasubject1 Advertiser
5  collect Location datasubject1 Advertiser
6  step
7  withdraw retro :consent1
```

[29] M. Klein, D. Fensel. "Ontology versioning on the semantic web." *Procedings of International Semantic Web Working Symposium*, pp. 75–91, 2001.

[30] J. Lin, et al. "Modeling users' mobile app privacy preferences: Restoring usability in a sea of permission settings." *Proceedings of 10th Symposium on Usable Privacy and Security*, pp. 199-212, 2014.

[31] H. J. Levesque, and R. J. Brachman. "Expressiveness and tractability in knowledge representation and reasoning 1." *Computational intelligence* 3.1, pp. 78-93, 1987.

[32] C. Lutz, F. Wolter, M. Zakharyaschev, "Temporal Description Logics: A Survey," *Proceedings of 15th International Symposium on Temporal Representations and Reasoning*, pp. 3-14, 2008.

[33] M. Nouwens, I. Liccardi, M. Veale, D. Karger, and L. Kagal, "Dark Patterns after the GDPR: Scraping Consent Pop-ups and Demonstrating their Influence," in Proceedings of the 2020 CHI Conference on Human Factors in Computing Systems, 2020.

[34] N. Noy, A. Rector, P. Hayes, C. Welty. "Defining N-ary Relations on the Semantic Web." *W3C W' Group Note*, 12, 2006.

[35] D. Nguyen, J. Park, & R. Sandhu. "Dependency Path Patterns as the Foundation of Access Control in Provenance-aware Systems," *Proceedings of 4th USENIX Conference on Theory and Practice of Provenance,* pp. 4, 2012.

[36] J. Park, and R. Sandhu. "Towards usage control models: beyond traditional access control." *Proceedings of 7th ACM Symposium on Access Control Models and Technologies,* pp. *57-64,* 2002.

[37] H. C. Pöhls, "Verifiable and revocable expression of consent to processing of aggregated personal data," in Lecture Notes in Computer Science (including subseries Lecture Notes in Artificial Intelligence and Lecture Notes in Bioinformatics), 2008, vol. 5308 LNCS, pp. 279–293.

[38] M. Robol, T. D. Breaux, E. Paja, P. Giorgini. "Consent verification under evolving privacy policies," *Proceedings of 27th IEEE International Requirements Engineering Conference,* pp. 422-427, 2019.

[39] M. D. Ryan, "Cloud Computing Privacy Concerns on Our Doorstep," *Communications of the ACM*, 54(1): 36-38, 2011.

[40] R. Sandhu, "Role-based Access Control," in *Advanced in Computers* 46: 237-286, 1998.

[41] R. Sandhu, E. Coyne, H. Feinstein, and C. Youman. "Role-Based Access Control Models". *Computer,* 29(2): 38–47, 1996.

[42] F. Schaub, R. Balebako, A. L. Durity, L. F. Cranor, "A design space for effective privacy notices," *Proceedings of Symposium on Usable Privacy and Security*, pp. 1-17, 2015.

[43] K. Slonneger, B.L. Kurtz, "Formal Syntax and Semantics of Programming Languages", Addison-Wesley, 1995.

[44] D. J. Solove. "A taxonomy of privacy," University of Pennsylvania Law Review, v. 154, n. 3, p. 477, 2006.

[45] S. Spiekermann, J. Grossklags, and B. Berendt. "E-privacy in 2nd generation E-commerce: privacy preferences versus actual behavior." *Proceedings of 3rd ACM Conference on Electronic Commerce*, pp. 38-47, 2001.

[46] M. Spiliopoulou, "Web usage mining for web site evaluation – making a site better fit its users," *Communications of the ACM*, 8(43): 127-134, 2000.

[47] M. Teltzrow, A. Kobsa. "Impacts of user privacy preferences on personalized systems." *Designing Personalized User Experiences in eCommerce*, pp. 315-332, 2004.

[48] R. K. Thomas, and R. Sandhu. "Task-based authorization controls (TBAC): A family of models for active and enterprise-oriented authorization management." *Database Security XI*. Springer, Boston, MA, 166-181, 1998.
32